\documentclass[sigconf,screen]{acmart}
\AtBeginDocument{%
  \providecommand\BibTeX{{%
    \normalfont B\kern-0.5em{\scshape i\kern-0.25em b}\kern-0.8em\TeX}}}

\setcopyright{acmcopyright}
\copyrightyear{2023}
\acmYear{2023}
\acmDOI{XXXXXXX.XXXXXXX}

\usepackage{ragged2e}
  \acmConference[CIKM '23]{32nd ACM International Conference on Information and Knowledge Management}{October 21--25, 2023}{University of Birmingham and Eastside Rooms, UK}
\acmBooktitle{CIKM '23: 32nd ACM International Conference on Information and Knowledge Management, October 21--25, 2023, University of Birmingham and Eastside Rooms, UK} 
\acmPrice{15.00}
\acmISBN{978-1-4503-XXXX-X/23/10}

\begin{document}


\title{ITA-ELECTION-2022: A multi-platform dataset of social media\\ conversations around the 2022 Italian general election}



\author{Francesco Pierri}
\email{francesco.pierri@polimi.it}
\affiliation{
  \institution{Dipartimento di Elettronica, Informazione e Bioingegneria, Politecnico di Milano}
  \city{Milano}
  \country{Italy}
}

\author{Geng Liu}
\email{geng.liu@polimi.it}
\affiliation{
  \institution{Dipartimento di Elettronica, Informazione e Bioingegneria, Politecnico di Milano}
  \city{Milano}
  \country{Italy}
}

\author{Stefano Ceri}
\email{stefano.ceri@polimi.it}
\affiliation{
  \institution{Dipartimento di Elettronica, Informazione e Bioingegneria, Politecnico di Milano}
  \city{Milano}
  \country{Italy}
}

\renewcommand{\shortauthors}{Pierri, Liu, and Ceri}
\begin{abstract}
Online social media play a major role in shaping public discourse and opinion, especially during political events.
We present the first public multi-platform dataset of Italian-language political conversations, focused on the 2022 Italian general election taking place on September 25th. Leveraging public APIs and a keyword-based search, we collected millions of posts published by users, pages and groups on Facebook, Instagram and Twitter, along with metadata of TikTok and YouTube videos shared on these platforms, over a period of four months. We augmented the dataset with a collection of political ads sponsored on Meta platforms, and a list of social media handles associated with political representatives. Our data resource will allow researchers and academics to further our understanding of the role of social media in the democratic process.
\end{abstract}







\maketitle

\section{Introduction}
Online social media provide researchers and academics with unprecedented opportunities to observe a wide range of political and societal phenomena \cite{rossi2021nearly}. They also play a critical role in shaping public opinion during political events \cite{vitak2011s}, and represent a rich source of data to study the interplay between political actors’ campaigns \cite{sahly2019social}, media outlets’ agenda settings \cite{kim2016first}, and users’ news consumption \cite{allcott2017social}. 

In Italy, as of 2022\footnote{\url{www.statista.com/statistics/1311549/top-social-platforms-italy/}}, YouTube is the platform used by the largest amount of internet users (88\%), followed by Meta platforms (64\%) and TikTok (54\%), whereas Twitter only accounts for approximately 7\%\footnote{\url{datareportal.com/reports/digital-2022-italy}}. However, previous studies of online social media during Italian elections and referendum mostly focused on Twitter \cite{rossi2021nearly}, due to the large availability of data via its APIs. In this work, instead, we present a public data resource of political conversations and user-generated content shared around the 2022 Italian general election, which allows researchers and academics to study multiple social platforms simultaneously.

The 2022 Italian general election was the first ever to take place in autumn, as a consequence of the fall of the government of national unity led by Mario Draghi in July\footnote{\url{https://www.reuters.com/world/europe/italys-meloni-sworn-head-right-wing-government-2022-10-22/}}. The election had a record-low voter turnout and it was won by the right-wing coalition of Giorgia Meloni with over 43\% of the vote share. Among the opponents, the Centre-left coalition led by Enrico Letta obtained approximately 25\% of the voters, the populist Movimento 5 Stelle led by former PM Giuseppe Conte reached less than 16\%, and the liberal and centrist Third Pole, which included former PM Matteo Renzi, obtained almost 8\% of the vote share.

We present \texttt{ITA-ELECTION-2022}, the first public multi-platform dataset of Italian-language political conversations taking place on online social media, with a focus on the 2022 Italian general election. We collected millions of social media posts from Facebook, Instagram and Twitter, as well as advertisements sponsored on Meta platforms and metadata for TikTok and YouTube videos shared on the aforementioned platforms. We finally augment the dataset with a collection of social media handles associated with Italian political representatives. To collect the data, we employed a snowball sampling procedure and curated a list of relevant terms to accordingly perform a keyword-based search during a period of four months (July 2022 - October 2022). We provide public access to the data via GitHub and DataVerse repositories, as detailed next. 


Given the nature of multi-platform, our dataset offers a unique opportunity for researchers to explore online political discourse during election seasons. It is particularly beneficial for researchers in those fields such as political science, sociology, digital ethics, and information disorder studies to study political communication, and information dissemination on social media.

The outline of this paper is the following: in the next section we review existing public data resources related to the present work; then, we describe the data collection procedure(s) carried out to build the dataset; next, we describe a few potential applications of the collected data; finally, we discuss limitations, draw conclusions and provide some ethical remarks.

\section{Related Work}
There are several public datasets that allow to study social media conversations around political issues. We focus our literature review on the Italian context, and then describe a few datasets related to other countries. We also refer the interested reader to \cite{rossi2021nearly} for an overview of studies that describe the interplay between social media and Italian politics.

\cite{valerio2018long} collect tweets in the Italian language continuously from 2012 to 2018, extracting a number of smaller datasets enriched with different kinds of annotations for linguistic purposes. They provide access to tweet IDs and annotations in a public repository.

\cite{PierriArtoni2020} analyze the prevalence of Italian disinformation spreading on Twitter in the five months preceding the 2019 European Parliament election. They collect over 300 k tweets sharing thousands of news articles originating from websites flagged as unreliable by journalists and fact-checkers, providing public access to tweet IDs and lists of websites. The same authors provide a similar dataset collected in a different period of 2019, and that contains tweets sharing links to mainstream and traditional news websites, both in the Italian and French language \cite{pierri2020diffusion}.

\cite{di2021content} study the polarization around the 2020 Italian constitutional referendum. They collect a dataset of 1.2 M tweets discussing the event -- and provide access to their IDs --, with the goal of designing a hashtag-based semi-automatic approach to label Twitter users' stance towards the referendum.

Following the COVID-19 pandemic, several researchers collected social media data to study conversations around the crisis, with a particular focus on the impact of vaccine misinformation. \cite{crupi2022echoes} study the evolution of Italian Twitter conversations around vaccines during the period 2019-2021,
whereas \cite{di2022vaccineu} collect tweets in multiple languages (French, German and Italian) during the first year of world vaccination programs. Both contributions give public access to tweet IDs, with the latter providing also a set of labeled pro/anti-vaccines tweets and hashtags that can be used for training machine learning classifiers.

\cite{calisir2020long} provide a dataset of tweets discussing Brexit for a period of 45 months, from January 2016 until September 2019. The data, which comprises 50.8 million tweets and 3.97 million users, is enriched with metadata such as the bot score of users, sentiment score of tweets, and political stance labels predicted by a classifier developed by the authors.

There is a large number of datasets that focus on the U.S. elections (both presidential and midterms), and we provide here a non-exhaustive list of available resources. \cite{hanna2011mapping} map candidates from the 2010 U.S. Midterm election with their Twitter accounts and a random sample of their followers. \cite{bovet2019influence}  collect over 171 M tweets in the English language, mentioning Donald Trump and Hillary Clinton during the 2016 U.S. Presidential election. \cite{deb2019perils} and \cite{yang2022twitter} collect tweets discussing the 2018 U.S. Midterm election, both using a hashtag-based search (e.g. tweets sharing the hashtag "\#ivoted" on election day) and querying Twitter APIs with general keywords related to the midterm election. \cite{chen2022election2020} provide a longitudinal dataset of over 1.2 billion U.S. politics- and election-related tweets shared around the period of the 2020 U.S. Presidential election. Related to the same election, \cite{abilov2021voterfraud2020} release a multi-modal dataset of 7.6 M tweets and 25.6 M retweets from 2.6 M users related to voter fraud claims. They augmented the data with cluster labels, users' suspension status, and perceptual hashes of tweeted images as well as aggregate data from external links and YouTube videos shared on Twitter. Lastly, \cite{aiyappa2023multi} recently released a multi-platform dataset of public social media posts related to the 2022 US Election, which is very similar to the present resource.

Unlike most existing datasets that focused only on Twitter or other individual platforms, our dataset covers a diverse set of social media platforms. In addition, we provide access to political ads placed on Meta plaforms as well as social media handles of political representatives. Insofar, our resource enable researchers to gain a more comprehensive understanding of social media discourse during the Italian election.

\section{Data Collection}

This section describes the data collection procedure(s) carried out to gather data from different social media platforms. We remark that we employed the same list of keywords related to the Italian election, which we obtained through a snowball sampling approach using Twitter data only, to query different APIs. Our dataset conforms with FAIR principles: it is \textit{Findable}, \textit{Accessible} and \textit{Reusable} as it is publicly accessible in an online Github\footnote{\url{github.com/frapierri/ita-election-2022}} and DataVerse repository\footnote{\url{doi.org/10.7910/DVN/EALXH2}}, where we provide the means to recreate it almost completely (see limitations discussed next). It is also \textit{Interoperable} as the data files are released in ``.csv" and ``.txt" formats.  We summarize some statistics of the dataset in Table \ref{tab:statistics}. 

We remark that, at the time of this writing, \textit{it is unclear whether Twitter APIs will be as easily accessible to researchers} as they were during the collection of this dataset, and this might affect future usage of this resource. We encourage interested researchers in reaching out to us in case they find difficulties in accessing Twitter data.

\subsection{Twitter}

We collected all tweets in the Italian language related to the election by using \texttt{tweepy} Python library to query Twitter v1.1 Filter streaming API endpoint\footnote{developer.twitter.com/en/docs/twitter-api/v1/tweets/filter-realtime/overview} in the period September 2nd, 2022 - October 20th, 2022. We also leveraged Twitter historical Search API v2 endpoint\footnote{developer.twitter.com/en/docs/twitter-api/tweets/search/introduction} to collect tweets retrospectively in the period July 1st, 2022 - September 1st, 2022. To query Twitter APIs we employed a snowball sampling approach, following existing work \cite{di2022vaccineu,deverna2021covaxxy}, and generated a list of relevant keywords starting with seed terms such as ``elezioni2022" and ``elezioni"(which means "elections" in Italian); the final list contains 62 keywords and it is available in the repository associated with this paper. A sample is provided in Table \ref{tab:keywords}. The total collection of tweets contains 19,087,594 tweets shared by 618,089 unique users. We remark that to abide by Twitter's terms of service we only share tweet IDs publicly. These can be ``re-hydrated" to retrieve tweet objects, with the exception of removed or protected tweets, by querying Twitter API directly or using tools like \texttt{Hydrator}\footnote{\url{github.com/DocNow/hydrator}} and \texttt{twarc}.\footnote{\url{github.com/DocNow/twarc}}


\begin{table}[!t]
\centering
\small
\begin{tabular}{ll}
\hline
\textbf{Platform} & \textbf{Statistics} \\
\hline
Twitter & 19,087,594 tweets \\  
 & 618,089 unique accounts \\

\hline
Facebook & 1,142,812 posts \\
& 445,461 unique accounts \\
\hline
Instagram & 68,078 posts \\
&5,274 unique accounts \\
\hline
Meta & 29,211 ads \\
& 3,750 unique sponsors \\
\hline
YouTube & 22,754 unique videos (Twitter) \\
 & 17,401 unique videos (FB) \\
 \hline
TikTok & 1,903 unique videos (Twitter) \\
 & 1,744 unique videos (FB) \\
\hline
\end{tabular}
\caption{Statistics of the dataset.}
\label{tab:statistics}
\end{table}

\begin{table}[!t]
\centering
\small
\begin{tabular}{lll}
elezioni & partito democratico & berlusconi \\ \hline
renzi & movimento 5 stelle & salvini \\ \hline
calenda & di maio & politiche2022 \\ \hline
meloni & elezioni2022 & conte  \\ \hline
\end{tabular}
\caption{A sample of Italian language keywords related to the 2022 election that were used to retrieve social media posts in our dataset.}
\label{tab:keywords}
\end{table}

\subsection{Facebook and Instagram posts}

We collected Facebook and Instagram data by employing CrowdTangle, a public tool owned and operated by Meta~\cite{crowdtangle} that allows retrieving posts shared by public pages and groups with a certain amount of followers or that were manually added by other researchers on the platform \footnote{\url{help.crowdtangle.com/en/articles/1140930-what-data-is-crowdtangle-tracking}}. We queried the \texttt{/posts/search} endpoint\footnote{\url{github.com/CrowdTangle/API/wiki/Search}} using the same list of keywords employed for collecting Twitter data. For each post, the API returns several attributes related to the post and the page or group (to which we refer as ``account" in the following) that published it; the full list of attributes is available in the official documentation\footnote{\url{github.com/CrowdTangle/API/wiki}}. We retained only posts in the Italian language by filtering on the \texttt{languageCode} parameter: the final dataset contains 1,142,812 Facebook posts, shared by 445,461 unique pages and groups and generating over 233 M interactions (shares, comments, reactions), and 68,078 Instagram posts, shared by 5,274 unique pages and generating over 97 M interactions (likes and comments). 
We provide access to the URLs and IDs of these posts in the repository associated with this paper\footnote{\url{github.com/CrowdTangle/API/wiki/Posts\#get-postid}}.

\subsection{TikTok and YouTube videos}

We augmented our dataset of social media posts by extracting metadata for TikTok and YouTube videos shared in Facebook and Twitter messages present in our dataset.  It is worth mentioning that no such links are shared in Instagram posts. For what concerns YouTube, we identified all external links to the platform and employed the official YouTube API\footnote{\url{developers.google.com/youtube/v3}} to extract video information such as the author, channel ID, video title, description, top 10 popular comments, etc. The resulting collection yields metadata for 22,754 unique YouTube videos shared on Twitter and 17,401 unique YouTube videos shared on Facebook. For what concerns TikTok, given the lack of an official API, we employed \texttt{pyktok} Python library\footnote{\url{github.com/dfreelon/pyktok}} to collect metadata about TikTok videos such as the title, description, length as well as information about the author of the video.
The resulting collection yields metadata for 1,903 unique TikTok videos shared on Twitter and 1,744 unique TikTok videos shared on Facebook. We provide full access to this metadata in the repository associated with the paper.

\subsection{Facebook and Instagram ads}


\begin{figure}[!t]
    \centering
    \begin{minipage}{0.48\linewidth}
        \centering
        \includegraphics[width=\linewidth]{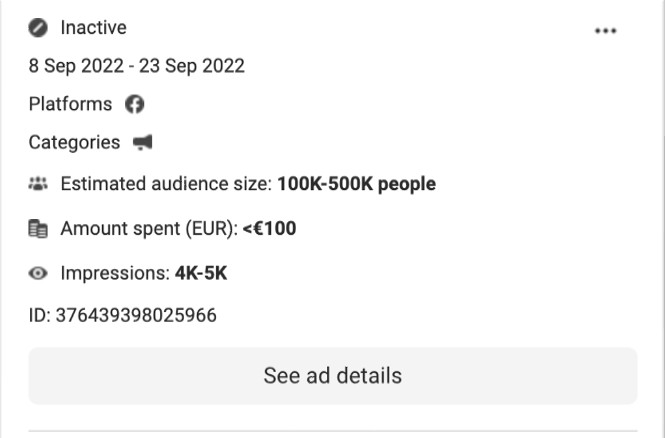}
    \end{minipage}
    \hfill
    \begin{minipage}{0.48\linewidth}
        \centering
        \includegraphics[width=\linewidth]{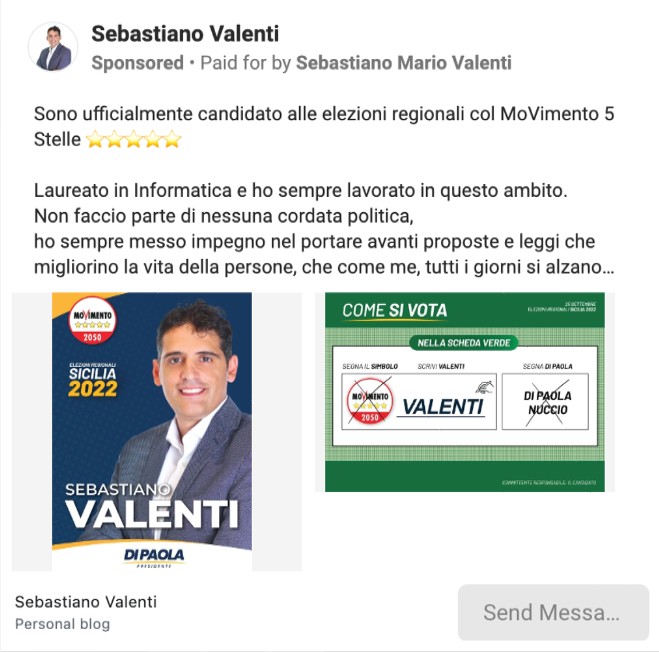}
    \end{minipage}
    \caption{Example of an ad run on Meta platforms along with the information provided by Meta Ad Library API.}
    \Description{Illustration of an advertisement on Meta platforms. The image shows the ad and corresponding data from the Meta Ad Library API.}
    \label{fig:ad-meta}
\end{figure}

We leveraged Meta Ad Library API\footnote{\url{www.facebook.com/ads/library/api}} to collect all ads about ``social issues, elections or politics" that were active on Meta platforms (i.e., Facebook, Instagram, Messenger, and the Audience Network) during the period from July 1st, 2022 - October 20th, 2022. We provide an example of a sponsored ad in Figure \ref{fig:ad-meta}. We queried the API with the same set of keywords mentioned beforehand; the API allows to search ads using one keyword at a time, and we queried the endpoint multiple times eventually discarding duplicated ads. The resulting collection contains 29,211 unique ads paid by 3,750 unique sponsors. For each ad, the API provides several different attributes: date of creation, period when the ad is active, name of the sponsor, message, platform on which the ad is active, lower and upper bound for the amount spent and the number of impressions generated, etc. Each of  these attributes is detailed in the API documentation, as referenced in the footnote above. In the repository associated with this dataset we provide access to the ID of ads, which can be then used to retrieve ads through Meta Ad Library interactive search console or API. In particular, to abide by Meta's terms of use, an identification procedure is required to access the API endpoint, whereas the interactive search console only requires a Meta account.

\subsection{Social media handles of political representatives}
We compiled a list of Facebook, Instagram and Twitter handles of elected members in the Senate and Chamber of deputies based on the official list released by the Italian Ministry of Interior\footnote{ \url{github.com/ondata/elezioni-politiche-2022}}. Specifically, for each representative, we manually checked whether their official account was present on the three platforms. 
In our dataset, we matched over 450 Twitter accounts and Facebook pages, and approximately 300 Instagram accounts.  The full list is available in the repository associated with this paper. We refer the interested reader to a similar useful resource presented by \cite{haman2021politicians}, who provide an online running database of politicians' activity on social media (currently only Twitter is supported) spanning multiple countries.

\section{Potential Applications}
There are several potential applications for our dataset, which can consider a single platform or multiple ones at the same time.

Interested researchers could further current understanding of polarization processes taking place during election seasons by analyzing content shared on multiple social platforms at once. They could study whether ``echo-chamber" effects take place on different platforms, highlighting similarities and differences in their formation process.

Other researchers might leverage the data in order to study how political candidates interacted with potential voters on social media platforms, thus analyzing in detail the political communication strategies put in place by different candidates. They could also investigate the presence of correlational effects between online signals and electoral outcomes, or detect the presence of toxic and hateful speech originating in specific communities of political supporters. 

Some researchers could investigate the presence of mis/disinformation and astroturfing campaigns taking place in the run-up to the election, studying patterns of similarities and differences among different platforms. They could also analyze how fringe and harmful content spreads across communities present on different platforms, and whether influential accounts play a role in amplifying certain malicious narratives.

\section{Conclusions}
We released \texttt{ITA-ELECTION-2022}, a large-scale dataset of social media posts in the Italian language discussing the 2022 Italian General election, which took place on 25th September 2022, spanning multiple online platforms and covering a period of four months. In addition to gathering posts shared on Twitter, Facebook and Instagram, we collected ads sponsored on Meta platforms, we extracted metadata for YouTube and TikTok videos shared on different platforms during the collection period, and we compiled a list of social media handles associated to political representatives that can be used to gather further data. We described in detail the collection procedures carried out to build the dataset, and we suggested potential directions for future research.

Our work is not without limitations. First, our keyword-based search might entail results that are not completely accurate, e.g., one of the terms employed for the query is ``conte", which might refer both to former PM Giuseppe Conte and football manager Antonio Conte. From another perspective, election-related terms might have been employed for marketing campaigns and promoting content that is not pertinent to the election. However, while we are unable to address these issues, which would require non-trivial efforts, researchers can further refine our data collection to meet their needs. Moreover, we performed a backward search to retrieve Twitter, Facebook and Instagram posts shared from July to September 2022, and we missed those that were deleted or removed during the same period. Similarly, by providing access only to IDs and URLs of collected posts, posts that have been removed or made private by users cannot be retrieved, thus limiting reproducibility analyses. Furthermore, we did not filter out the activity of automated and inauthentic accounts that might have polluted organic conversations around the election. Another limitation concerns Meta which, as highlighted in \cite{le2022audit}, might not accurately label all political ads as such and our collection might be missing some data. Finally, the user base of different platforms analyzed in this work might not be fully representative of the actual Italian population, and this should be taken into consideration by future research.

Despite these limitations, we believe that our dataset provides fertile ground for a number of intriguing and interesting research applications. We hope that this resource can advance our understanding of the interplay between online social media and democratic processes.

\section{Ethical considerations}
We performed our data collection and public release in complete agreement with the platforms' terms of service. We acknowledge that TikTok metadata was scraped from the platform, thus potentially violating the platform's terms of service, but this was due to the lack of an official API \cite{freelon2018computational}.  We do not directly share the raw content of social media posts, but rather provide access to IDs and URLs that can be used to retrieve the original data, with the exception of posts that have been deleted by platforms, and removed or made private by their author. We did not cause any harm to nor expose information about individual users in the process of collecting and releasing the data, with the only exception of political representatives. We understand that disclosing their social media accounts might open up to potential abuse by malicious actors, but at the same time, it enables researchers, journalists and other stakeholders to put important public actors, such as the members of the Italian Parliament and Senate, to scrutiny in order to better understand the influence of social media platforms on the democratic process. Finally, we reckon that Twitter might limit access to its public APIs, and we encourage interested researchers in contacting us may they encounter difficulties in retrieving Twitter data.


\bibliographystyle{ACM-Reference-Format}
\bibliography{bib}

\appendix

\end{document}